\documentclass[fleqn,twoside]{article}
\usepackage{espcrc2}

\usepackage{graphicx}
\usepackage[figuresright]{rotating}

\newcommand{\AmS}{{\protect\the\textfont2
  A\kern-.1667em\lower.5ex\hbox{M}\kern-.125emS}}

\title{Benchmarking and tuning the MILC code on clusters and supercomputers}

\author{Steven Gottlieb\address{Department of Physics---SW117; Indiana
	University; Bloomington, IN 47405; USA}
	\address{Theory Group MS106; Fermilab; P.O. Box 500;
	Batavia, IL 60510--0500; USA}
        \thanks{At Fermilab until June 15, 2002.}
        }
       
\begin{document}

\begin{abstract}
Recently, we have benchmarked and tuned
the MILC code on a number of architectures including Intel Itanium and
Pentium IV (PIV), dual-CPU Athlon, and the latest Compaq Alpha nodes.
Results will be presented
for many of these, and we shall discuss some simple code changes that can
result in a very dramatic speedup of the KS conjugate gradient on processors
with more advanced memory systems such as PIV, IBM SP and Alpha.
\vspace{1pc}
\end{abstract}

\maketitle

\section{INTRODUCTION}
This contribution is a condensation of a 16 page poster with 17 tables
of benchmarks.  The poster is available on the web \cite{posterURL}.

Benchmarks presented here are
for the Conjugate Gradient algorithm with Kogut-Susskind quarks,
not just for $D\mkern-12mu\slash$.
They are done within the context of a complete application for creation of 
gauge fields using the $R$-algorithm \cite{algorithm}.
The application uses even-odd checkerboarding, which reduces possible reuse
of data in cache.
Even the single CPU benchmarks are done with a
fully parallel application that splits the computation within 
$D\mkern-12mu\slash$ into
two stages to accommodate the need to wait for boundary values that would
come from another node in a multiCPU run.
This also reduces potential cache reusage.
On some of the architectures, we make use of assembly code for 
basic $SU(3)$ arithmetic routines or for prefetching data to cache.
We use Kogut-Susskind quarks
for benchmarking because they are used in our dynamical quark
calculations.
KS quarks are more demanding than Wilson quarks in terms of memory bandwidth.
In single precision, the former require 1.45 bytes/flop of input data and
produce 0.36 byte/flop of output.  
For Wilson quarks only 0.91 bytes/flop of input is
required and output is unchanged.
Thus, it should not be surprising to find that a Wilson quark code can
achieve higher speed than reported here \cite{luscher}.

\section{ARCHITECTURES}

Since August
2000, MILC has been working with Intel and NCSA under a non-disclosure
agreement to tune our code for the Itanium processor.
In December 2000, we were allowed to report first results without assembly
code\cite{ccp2k}.
Some limited results with assembly code were reported at Linux World last
January.  We may now talk more freely about results on Itanium.

MILC has had several months of
production running on the initial Terascale Computer System at the Pittsburgh
Supercomputer Center.
It is based on Compaq ES40 nodes that contain 667 MHz EV67 Alpha chips.
The full 6 TF computer will be based on 1000 MHz EV68 chips.
At the end of March, we were given access to the first 
ES45 node at PSC that contains that chip.

IBM SP tests have been run on either the
Indiana University SP or Blue Horizon at SDSC.
They have 375 MHz Power 3 chips deployed on 4-way and 8-way SMP nodes,
respectively.

During the Spring, we had access to a 1.5 GHz Pentium IV system
and a dual 1.2 GHz Athlon system, thanks to NCSA and Penguin Computing, 
respectively.

\section{CODE CHANGES}
The work on the Itanium processor
was carried out in conjunction with two Intel engineers, Gautham Doshi and
Brian Nickerson.
Doshi worked on in-lining and optimizing compiler flags for the C code.
Nickerson wrote assembly code that includes prefetching and 
looping over sites.  The changes described next have not yet
been tried on Itanium.

The MILC code data structure is ``site major,'' 
{\it i.e.}, there is a structure for each site that contains
all the physical variables for that site. 
The lattice is an array of site structures.
Adding variables to the application is quite easy:
One only needs to modify the site
structure, and when the lattice is allocated, the new variables
will be globally accessible.  This Spring, we tested performance enhancements
from temporary allocations of ``field major'' variables for the conjugate
gradient routine.
On chips with wider cache lines, this results in substantial speedups.
The gauge fields and necessary vectors are
copied to temporary variables that are much better localized in memory.
If a cache line contains data not needed for the current site, it is most
likely the data required for the next site to be computed, rather than a
different physical variable, as would be found in the next bytes of
the site structure.  I suggested these changes to Dick Foster, of Compaq,
who implemented them and improved prefetching.

\section{SINGLE NODE RESULTS}
The benchmarks presented here were run on lattices of size $L^4$.
They are all for single precision gauge links and vectors, with dot products
accumulated in double precision.
The fermion matrix is either for the
Kogut-Susskind (KS) or fat-link plus Naik (fat-Naik) action \cite{naik}.
For production runs, we are using the ``Asqtad'' action \cite{asqtad}.
(The performance of the inverter is independent of the details of the
fattening).

\subsection{Itanium}
Results on Itanium without assembly code were presented at CCP2000 \cite{ccp2k},
and are available on the web. With an 800 MHz processor, performance was
916, 867 and 732 MF for $L=4$, 6 and 8, respectively.  Because of memory
access issues, performance drops to 326 MF for $L=14$.  
With Nickerson's assembly
code, the numbers are quite impressive.  We have 1223, 1139 and 938 MF
for $L=4$, 6 and 8, respectively, and even for $L=14$, we achieve 464 MF.
The field major code has not yet been tried on Itanium.

\subsection{Alpha}
In Table~\ref{tab:alpha}, 
we compare the performance of the old site major code with
the new field major code.  We present results for both the 667 MHz EV67
chips in the ES40 and the 1000 MHz EV68 chips in the ES45.  We can see
substantial speedups both from the newer processor and the code
improvement.  Currently, Itanium is the performance leader for smaller
$L$, while Alpha leads for large $L$.  Of course, the codes are different
and considerable work would be required to combine both the benefits of
assembly code (with loop control) and field major organization on
each chip.

\subsection{Power 3}

The IBM SP really benefits from the new field major code.  Table~\ref{tab:sp}
shows
the performance and speedup for various $L$.  The substantial falloff
with increasing $L$ has been greatly ameliorated, and the overall
performance level has increased substantially even for small $L$.
These results and the corresponding multinode results were obtained on the
Indiana University SP.
Now let's turn to the commodity processors.

\begin{table}[th]
\caption{Megaflop rate on Alpha Processors}
\label{tab:alpha}
\begin{tabular}{cccc}
\hline 
L & ES40 & ES45 & ES45\\
 & site major & site major & field major\\
\hline 
6 & 517 & 731 & 977 \\
8 & 495 & 701 & 843 \\
10 & 395 & 548 & 934 \\
12 & 249 & 395 & 778 \\
14 & 253 & 347 & 609 \\
\end{tabular}
\medskip
\caption{Megaflop rate and speedup on IBM SP}
\label{tab:sp}
\begin{tabular}{cccc}
\hline 
L & site major & field major & speedup\\
\hline 
4 & 512 & 663 & 1.29 \\
6 & 458 & 705 & 1.54 \\
8 & 391 & 682 & 1.74 \\
10 & 215 & 557 & 2.58 \\
12 &  158 & 528 & 3.35 \\
14 & 135 & 449 & 3.32 \\
\end{tabular}
\end{table}
\begin{table}[ht]
\caption{Megaflop rate and speedup on 1.5 GHz PIV}
\label{tab:piv}
\begin{tabular}{cccc}
\hline 
L & site major & field major & speedup\\
\hline 
4 & 591 & 577 & 0.98 \\
6 & 240 & 503 & 2.10 \\
8 & 220 & 481 & 2.19 \\
10 & 208 & 491 & 2.36 \\
12 & 205 & 480 & 2.34 \\
14 & 202 & 469 & 2.33 \\
\end{tabular}
\medskip
\caption{Megaflop rate per CPU on dual 1.2 GHz Athlon MP system}
\label{tab:athlon}
\begin{tabular}{ccccc}
\hline 
L & site m. & site m. & field m. & field m.\\
 & single & dual & single & dual\\
\hline 
4 & 590 & 464 & 654 & 457 \\
6 & 203 & 167 & 336 & 251 \\
8 & 176 & 142 & 298 & 232 \\
10 & 170 & 134 & 289 & 228 \\
12 & 165 & 132 & 287 & 239 \\
14 & 166 & 133 & 281 & 218 \\
\end{tabular}
\end{table}

\begin{table}[bht]
\caption{Megaflop rate for $8^4$ sites per CPU}
\label{tab:summary}
\begin{tabular}{lcccccc}
\hline
& 1 & 4 & 128 & 256 \\
\hline
ES45 (field) & 839 & 621 \\
ES40 (site) & 495 & 425 & 302 & 262 \\
SP (site) & 375 & 340 & 204 & 181 \\
SP (field) & 624 & 529 & 176 & 140 \\
Itanium (site) & 503 & 304 \\
Platinum (site) & 139 & 94 & 75 \\
Platinum (field) & 159 & 107 & 71 \\
Scali (field) & 72 & 63 \\
\end{tabular}
\end{table}

\subsection{Intel IA32 and AMD}
Both Pentium IV and AMD Athlon MP processors show excellent speedup
on the new field major code.  Details appear in Tables \ref{tab:piv} and 
\ref{tab:athlon}.
For the Athlon we had a dual CPU system and show results for both
one and two processors.  The Pentium IV is performing at almost 500 MF
even for $L=8$ and greater.  It is not as fast as the previous chips 
discussed, but it is certainly very cost effective.  The Athlon system
has DDR memory rather than Rambus (RDRAM).  
One can see that for $L=4$ for which the problem fits in cache,
the Athlon, despite its slower clock speed out performs the Penium IV.
However, for larger $L$, access to memory becomes crucial and the
Pentium IV excels.  It would be interesting to try a Pentium IV motherboard
that uses DDR memory.
On the dual Athlon system the Fat-Naik inverter was benchmarked and found
to be 10--20 MF faster than KS \cite{posterURL}.

\section{MULTINODE RESULTS}

The program Netpipe has been used to compare message passing speeds of
Fast Ethernet, Myrinet, Scali, Quadrics and the IBM SP network.
Fast Ethernet only achieves 20--60 Mbit/s for messages of the size needed
during the conjugate gradient (800--30K bytes).  The other networks,
except for Quadrics are about a factor of 10 faster.  Quadrics is about
an additional factor of two faster.

Tables of results are available \cite{posterURL} for ES45 with up to
four CPUs, the ES40 with up to 256 CPUs, the IBM SP with up to 256 CPUs,
the prototype Itanium cluster with up to 16 CPUs, the Platinum (Pentium III)
cluster with up to 128 CPUs and a Pentium II cluster with Scali interconnect.
Here we just display results for $L=8$.  The table indicates whether the
code was site major or field major.  Scali is limited by the power of the
CPU.  Results on larger numbers of ES45 and Itanium nodes should be available
in late October.

Thanks to Compaq, Intel, NCSA, Penguin Computing, PSC, SDSC, UITS and the
MILC Collaboration.


\end{document}